\documentclass[12pt,preprint]{aastex}
\usepackage{graphicx}
\begin{document}

\slugcomment{accepted, Astrophysical Journal}

\title{Water, Methane, and Carbon Dioxide Present in the Dayside
  Spectrum of the Exoplanet HD~209458b}

\author{M.~R. Swain\altaffilmark{1},
G. Tinetti\altaffilmark{2},
G. Vasisht\altaffilmark{1}, 
P. Deroo\altaffilmark{1},
C. Griffith\altaffilmark{3},
J. Bouwman\altaffilmark{4},
Pin Chen\altaffilmark{1},
Y. Yung\altaffilmark{5}, 
A. Burrows\altaffilmark{6},
L.~R. Brown\altaffilmark{1},
J. Matthews\altaffilmark{7},
J.~F. Roe\altaffilmark{9},
R. Kuschnig\altaffilmark{8}, \&
D. Angerhausen\altaffilmark{10}}

\altaffiltext{1}{Jet Propulsion Laboratory, California Institute of
Technology, 4800 Oak Grove Drive, Pasadena, CA 91109}

\altaffiltext{2}{University College London, Gower Street, London WC1E 6BT, UK}

\altaffiltext{3}{University of Arizona, Lunar and Planetary
  Laboratory, Space Science Bldg. Room 525, 1629 E. University
  Bvld. Tucson, AZ 85721}

\altaffiltext{4}{Max-Planck Institute for Astronomy, Konigstuhl 17,
  D-69117 Heidelberg, Germany}

\altaffiltext{5}{Division of Geological and Planetary Sciences,
  California Institute of Technology, Pasadena, CA 91125}

\altaffiltext{6}{Dept. of Astrophysical Sciences, Princeton
  University, Princeton, NJ 08544}

\altaffiltext{7}{Dept. of Physics \& Astronomy, University of British
  Columbia, Vancouver, BC V6T 171 Canada}

\altaffiltext{8}{NASA Ames Research Center, MS 244-30, Moffett Field, CA 94035}

\altaffiltext{9}{Institut fur Astronomie, Universitat Wien
Turkenschanzstrasse 17, A-1180 Wien, Austria}

\altaffiltext{10}{Institute of Space Systems, University Stuttgart;
DAAD-Fellow}

\begin{abstract}

Using the NICMOS instrument on the Hubble Space Telescope, we have
measured the dayside spectrum of HD~209458b between 1.5--2.5
$\mu$m. The emergent spectrum is dominated by features due to the
presence of methane (CH$_{4}$) and water vapor (H$_{2}$O), with
smaller contributions from carbon dioxide (CO$_{2}$).  Combining this
near-infrared spectrum with existing mid-infrared measurements shows
the existence of a temperature inversion and confirms the
interpretation of previous photometry measurements.  We find a family
of plausible solutions for the molecular abundance and detailed
temperature profile.  Observationally resolving the ambiguity between
abundance and temperature requires either (1) improved wavelength
coverage or spectral resolution of the dayside emission spectrum, or
(2) a transmission spectrum where abundance determinations are less
sensitive to the temperature structure.

\end{abstract}


\section{Introduction}

HD~209458b has the distinction of being the first identified
transiting extrasolar planet \citep{charbonneau00}.  This 0.69
Jupiter-mass planet orbits a G0V stellar primary (K$_{mag}$=6.3) every
3.52 days and was the first exoplanet with a detected atmosphere
\citep{charbonneau02}.  It was also one of the first planets found to
emit detectable amounts of infrared radiation \citep{deming05}.  The
dayside mid-infrared spectrum for this planet, based on Spitzer IRS
observations, has been previously obtained
\citep{richardson07,swain08a}. Mid-infrared dayside photometry
\citep{knutson08} provides evidence for an atmospheric temperature
inversion \citep{burrows07} and has been used to support a
classification scheme for the hot-Jupiter type exoplanets
\citep{fortney08}.  Mid-infrared photometric measurements of the
orbital phase light curve of HD~209458b \citep{cowan07} find that the
dayside and nightside brightness temperatures are similar.  Models of
atmospheric circulation for HD~209458b suggest that the zonal wind
structure is a function of altitude \citep{showman08}, and
mid-infrared dayside emission could vary
\citep{rauscher07,rauscher08,showman08}.

The recent announcement of the detection of H$_{2}$O and CH$_{4}$ in
the terminator regions of the hot-Jupiter HD~189733b (Swain, Vasisht
\& Tinetti 2008b; hereafter SVT08) via near-infrared spectroscopy
using the Hubble Space Telescope (HST) demonstrated that molecular
spectroscopy of bright transiting planets is possible.  The subsequent
HST detection of H$_{2}$O, CO$_{2}$, and carbon monoxide (CO) on the
dayside of HD 189733b provides significant constraints for atmospheric
models and suggests that exoplanet molecular spectroscopy may become
routine (Swain et al. 2009; hereafter S09) in the near future.  This
is significant because molecules provide a powerful tool for
determining the conditions, composition, and chemistry of exoplanet
atmospheres.  In this paper, we report on the first ever spectroscopic
detection of molecules in the atmosphere of HD~209458b, making this
the second exoplanet for which molecular spectroscopy has now been
demonstrated.

\section{Observations \& Data Analysis}

We observed HD~209458 for five consecutive HST orbits using the NICMOS
camera in imaging-spectroscopy mode covering the wavelengths 1.5--2.5
$\mu$m using the G206 grism.  As in previous observations (SVT08,
S09), the DEFOCUS mode was used, resulting in a spectral resolution of
$R \simeq 40$.  Observations were conducted on June 15, 2008, from UT
09:38:59 to UT 16:45:24. A total of 310 usable snapshot spectra were
acquired with an individual exposure time of $T = 4.06$ s. Using the
system ephemeris \citep{knutson07} we determine that the bulk of the
first two orbits ($O_1$ \& $O_2$) coincides with the pre-ingress light
curve, the third and part of the fourth orbit ($O_3$ \& $O_4$) cover
the occultation, while the fifth orbit ($O_5$) is post-egress (see
Fig. 1). We find that 75 exposures cover the full eclipse, 43
exposures are in ingress/egress, and 192 exposures cover the
pre-ingress or post-egress intervals. The effective exposure time for
each spectrum was $T = 4.06$ s.  To determine the spectrometer
wavelength calibration, narrow band (F190) filter calibration
exposures were acquired during $O_1$.

Our data analysis methods, based on decorrelation of instrument
parameters from the measured spectrophotometric light curves, have
been described previously (SVT08, S09).  For the results presented
here, the data-analysis approach remains largely unchanged but is
updated in two significant ways.  First, the decorrelation for
instrument parameters is based on detector pixel values; in the past
we averaged the detector data to the defocused spectral resolution
prior to the decorrelation.  The new approach has the advantage of
determining the corrected, decorrelated spectrum using the full
oversampling of the instrument.  Second, we now derotate and shift the
spectrum (via interpolation) to achieve the same sampling at all
wavelengths and times.  This effectively places observations in a
common, pixel-based, reference frame.  We tested this new reduction
method by computing a dayside spectrum of HD~189733b and comparing it
to our previous result (S09); the agreement between the two methods
was found to be excellent, with differences between the spectra always
within the statistical uncertainty of the measurement.  As with our
previous observations of the primary and secondary eclipse of
HD~189733b, we found the instrument parameters for the first orbit,
$O_{1}$, differed significantly from the remaining orbits,
$O_{2}$--$O_{5}$; following our previous approach (SVT08, S09), we
omitted the $O_{1}$ spectra from the analysis.  To add an additional
constraint to any possible orbit-to-orbit brightness changes, we
arranged for contemporaneous, high-precision, visible photometry using
the Micro-variability and Oscillations of Stars (MOST) satellite
\citep{walker03}.  The MOST observations were timed to provide a
photometry measurement during each HST orbit (see Fig. 1), and the
data were calibrated using the method described by \cite{rowe08}.

\section{Discussion}

The near-infrared dayside spectrum of HD~209458b shows three clear
peaks at 1.6, 1.8--1.9, and 2.2 $\mu$m, while between the peaks the
flux density decreases to near-zero (within the measurement error).
The spectrum has an average uncertainty per data point of $\sim
7.5\times10^{-5}$, corresponding to a dynamic range of 13,300:1 (75
ppm), with a wavelength resolution of $\frac{\lambda}{\Delta \lambda} =
35$ at 2 $\mu$m.  To maximize wavelength coverage for the spectral
interpretation, we incorporated mid-infrared Spitzer photometry data
\citep{knutson08} and spectroscopy data (originally obtained by
Richardson et al. 2007).  We have previously undertaken our own
calibration of these Spitzer IRS data \citep{swain08a},
and we use those results in the discussion that follows.

The interpretation of the data was done using two independent spectral
retrieval methods incorporating a line-by-line radiative transfer model
(\cite{griffith98,tinetti07a,tinetti07b}, SVT08, S09), and the results
from both models are in agreement.  The radiative transfer
calculations assume local thermal equilibrium (LTE) conditions, as
expected for pressures exceeding 10$^{-3}$ bar that are probed by the
near-infrared spectrum. We evaluated a variety of temperature ($T$)
together with the effects of CH$_{4}$, H$_{2}$O and CO$_{2}$, which
are assumed to have constant mixing ratios. We use the CH$_{4}$ line
list of \cite{nassar03} for temperatures at 800, 1000, and 1273 K to
quantify the affects of CH$_{4}$ at wavelengths of 1.58--2.5 $\mu$m,
outside of which we use absorption parameters by \cite{rotham05}. The
hot water lines from \cite{barber06} and \cite{zobov08} were used to
quantify the water features. Absorption by CO$_{2}$ is calculated from
the HITRAN hot CO$_{2}$ line list \citep{tashkun03}. Absorption
coefficients of H$_{2}$O and CO$_{2}$ are calculated using
line-by-line techniques every 0.004 cm$^{-1}$ wavenumbers. The
radiative transfer calculation has 80 vertical grid points that extend
from 10$^{-7}$ to 10 bar. The effects of particulates are excluded.

The combination of near-infrared and mid-infrared data probes the
temperature profile over a pressure range of a few bars to 10$^{-5}$
bars and provides evidence for a temperature inversion.  The 8 $\mu$m
brightness temperature is 1770 K, while the 2.2--2.4 $\mu$m brightness
temperature ranges from 1100--1600 K; this temperature difference,
combined with the difference in pressure probed by these wavelengths,
indicates the presence of a temperature inversion.  We find that a
temperature inversion could occur somewhere between $\sim$
$10^{-1}$--$10^{-4}$ bar, where temperatures increase by roughly
500--700 K.  This temperature-inverted region resembles those of
planetary stratospheres; it occurs at a similar pressure level and
causes a convectively stable temperature profile.  The model spectrum
from three exemplar temperature profiles (inverted) are shown in Fig. 3
(together with a temperature profile with no inversion), and the model
residuals (Fig. 4) show the improved fit with a temperature inversion.
The presence of the stratosphere coupled with a relatively steep
stratospheric temperature gradient implies the existence of a local
absorber that heats the atmosphere and maintains the local temperature
structure.  The temperature profile we derive confirms the previous
identification of a $T$ inversion \citep{burrows07,knutson08}, but the
profile reported here differs in terms of the detailed shape.

Our estimates for the temperature profile and molecular mixing ratios
are based on modeling the combined near-infrared and mid-infrared data
sets.  However, the recent finding of variability in the mid-infrared
emission spectrum of HD~189733b \citep{grillmair08} suggests caution
is needed when deriving constraints from the noncontemporaneous mid-IR
data.  Although the presence of H$_{2}$O and CH$_{4}$ can explain most
of the spectrum, we find that an additional absorber is required
around 2.1 $\mu$m, and we attribute this to CO$_{2}$.  The evidence
for a temperature inversion comes from the mid-infrared spectrum, and
the mid-infrared data provide some additional constraints on the
mixing ratios for H$_{2}$O and CO$_{2}$.  We find that a range of
temperature profiles and molecular mixing ratios are consistent with
the data as illustrated by the models shown in Fig. 4 and the sample
contribution functions shown in Fig. 5, which correspond to three
possible temperature profiles.  Providing improved observational
constraints on the location of the tropopause and the molecular mixing
ratios will require additional observations, for which there are two
possibilities.  The first is to improve the spectral coverage and/or
spectral resolution of the dayside spectrum, which could be done in
the future with JWST.  The second is to use a transmission spectrum
(obtained during primary eclipse) to estimate the molecular abundances
in the terminator regions; subject to some assumptions, this could be
used to place constraints on the abundance of one or more molecules.
HST transmission spectra for this planet have been obtained and are
being analyzed by our team.  However,the interpretation of the
transmission spectrum is complex due to the presence of temporal
variability (the subject of a forthcoming paper).

There are two important caveats concerning our derived temperature
profiles.  First is the assumption of emission from a uniform disk
rather than from a realistic irradiated hemisphere; as such, the
spectrum and T-profile we derive is disk-averaged.  Second, the
spectral retrieval process used in our best fitting models implicitly
acknowledges the possible role of dynamics in establishing atmospheric
temperature by relaxing typical 1-line model constraints on the
temperature gradient.  Thus, portions of the atmosphere of HD~209458b
may support departures from radiative equilibrium.  Fully
self-consistent modeling would require handling heat advection in the
context of a global circulation model, which is beyond the scope of
this paper.  It is worth noting that the presence of a
relatively strong dayside temperature inversion significantly
complicates the spectral retrieval process for HD~209458b relative to
the case of HD~189733b.

Given that the dayside emission spectra for HD~189733b and HD~209458b
have been observed with nearly identical instrument configurations, we
can make a preliminary comparison of these two planets (see Fig. 6).
For the present, we restrict this discussion to the near-infrared to
avoid the complications introduced by the presence of mid-infrared
variability observed in HD~189733b \citep{grillmair08}.  In both
planets, the near-infrared dayside emission spectrum probes the $5
\leq P \leq 10^{-2}$ bar portion of the atmosphere.  Similarities --
both planets show the presence of H$_{2}$O, CO$_{2}$, and
$\frac{dP}{dT}\leq 0$ for pressures near 1 bar.  Differences -- the
abundance of CH$_{4}$ is significantly enhanced in HD~209458b relative
to HD~189733b, and the temperature at 1 bar is higher in HD~209458b,
as well.  The observed differences in the near-infrared dayside
spectra of these two hot-Jupiters are likely due to differences in
temperature and composition; HD209458b is dominated by CH$_{4}$
absorption features, while HD~189733b is dominated by absorption from
H$_{2}$O and CO$_{2}$.

\section{Conclusions}

In summary, we have presented the first near-infrared spectrum of
light emitted by HD~209458b. Using an iterative forward model approach
for spectral retrieval, we find that H$_{2}$O, CH$_{4}$, and CO$_{2}$,
together with a temperature inversion, are present in the dayside
atmosphere of HD~209458b.  There are a range of temperature profiles
and molecular abundance solutions that are consistent with the data.
Additional observational constraints on the atmospheric temperature
structure and composition will require either improved wavelength
coverage/spectral resolution for the dayside spectrum or a
transmission spectrum.  We note that some of the temperature profiles
consistent for these observations raise the question of whether the
dayside atmosphere is in radiative equilibrium.  Although advection of
heat could support departures from radiative equilibrium, our present
knowledge of most molecular opacities at high temperatures limits our
ability to determine decisively whether the radiative equilibrium
condition is met or not; thus there is an urgent need for further
laboratory studies to obtain molecular databases for determining high
temperature opacities of the most common molecules expected in
hot-Jupiter atmospheres.

\acknowledgments

We appreciate the Director's time award for these observations, and we
thank Tommy Wiklind, Beth Padillo, and other members of the Space
Telescope Science Institute staff for assistance in planning the
observations.  We also thank Jonathan Tennyson and Bob Barber for help
with the water line list.  G. Tinetti was supported by the Royal
Society.  A portion of the research described in this paper was
carried out at the Jet Propulsion Laboratory, under a contact with the
National Aeronautics and Space Administration.

\begin{figure}[h!]
\begin{center}
\epsscale{0.5}
\includegraphics[angle=0,scale=0.9]{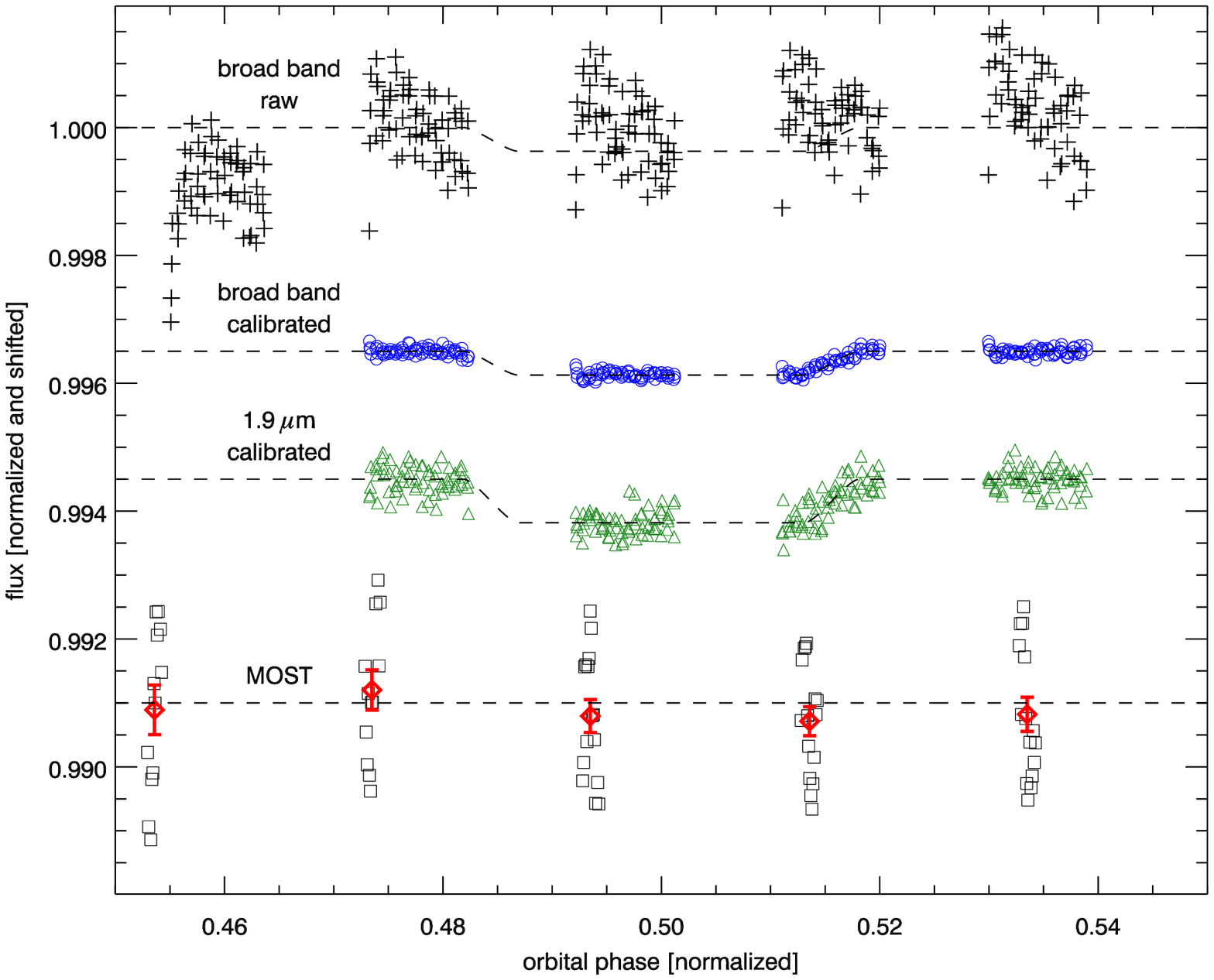}
\caption{This figure shows a set of normalized light curves plotted as
a function of orbital phase.  Gaps in the data are unavoidable because
HD~209458 is not in the continuous viewing zone of Hubble.  The top
curve is the raw, broad-band, NICMOS data showing large internal
scatter and a systematic offset affecting the first orbit.  The second
curve shows the calibrated, broad-band, NICMOS data (we exclude the
first orbit); a shallow eclipse is visible.  The third curve shows the
calibrated 1.9 $\mu$m data.  The larger eclipse depth at this
wavelength (compared to the calibrated broad-band data) is readily
apparent.  The fourth curve shows the contemporaneous MOST visible
photometry data.  Both the individual and averaged data are shown.
The MOST data are consistent with no changes in the system brightness
at visible wavelengths.
\label{fig:lightCurve}}
\end{center}
\end{figure}

\begin{figure}[h!]
\begin{center}
\epsscale{0.5}
\includegraphics[angle=0,scale=1.0]{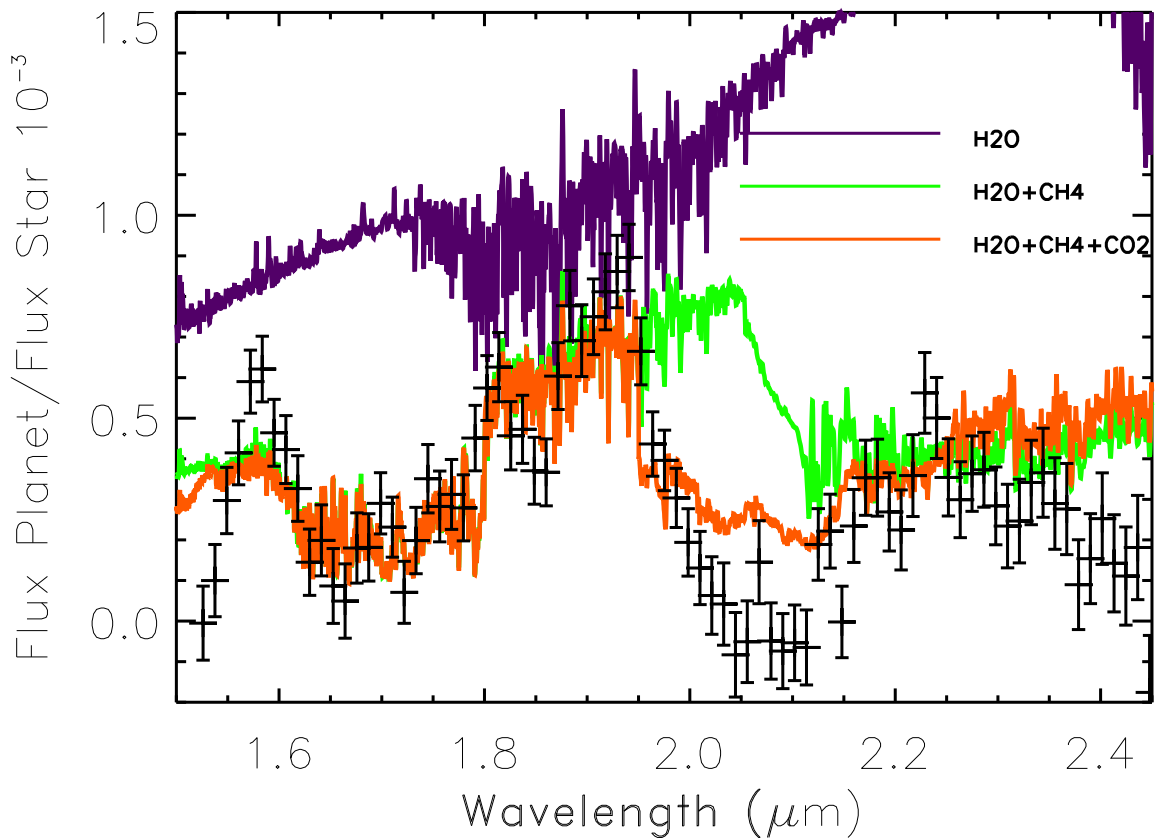}
\caption{A comparison of the observed spectrum (black points with
  $\pm$ 1-$\sigma$ error bars) to a sequence of models showing the
  affect of H$_{2}$O, CH$_{4}$, and CO$_{2}$.  This comparison shows
  the portion of the near-infrared spectrum where each molecule makes
  a significant contribution.  CH$_{4}$ and H$_{2}$O explain most of
  the spectral features, although additional absorption is needed
  around 2.0 $\mu$m (provided by CO$_{2}$).  Here, the observed
  spectrum has been smoothed with a 5-point scrolling boxcar and thus
  adjacent data points are not statistically independent.
\label{fig:molecules}}
\end{center}
\end{figure}

\begin{figure}[h!]
\begin{center}
\epsscale{0.5}
\includegraphics[angle=0,scale=1.0]{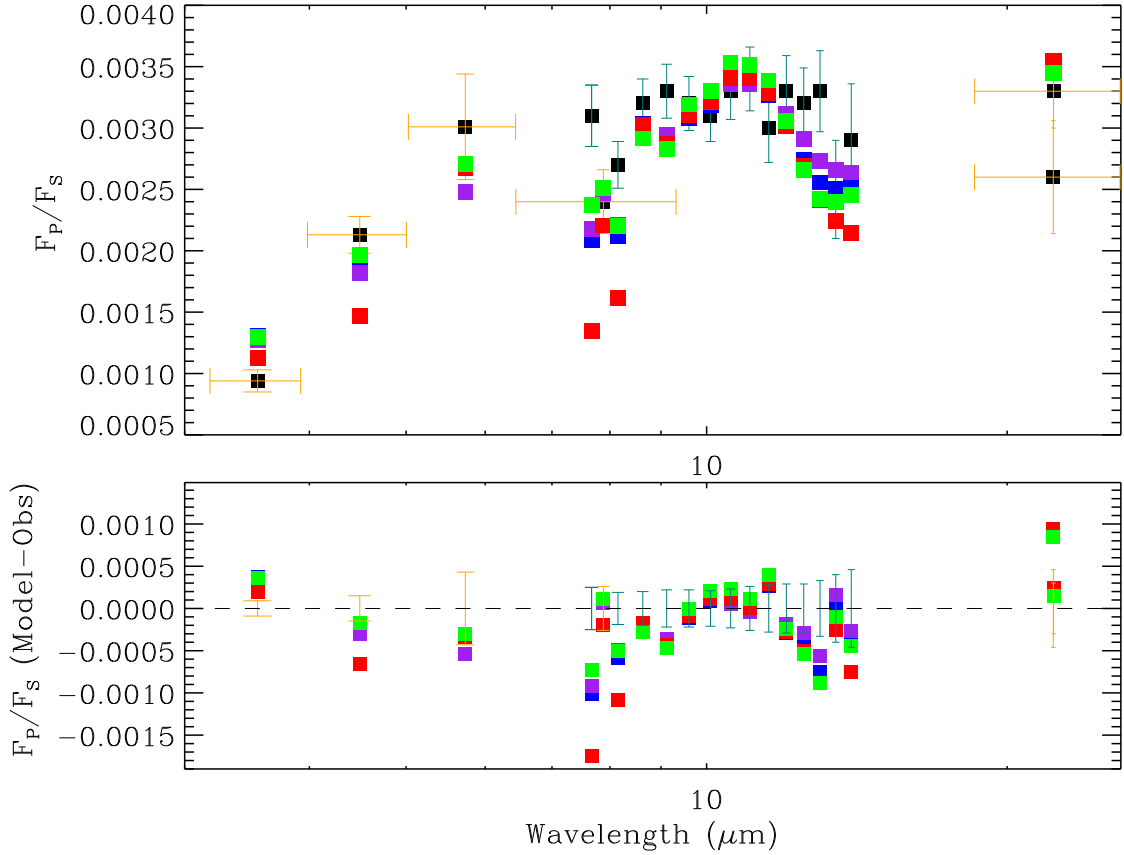}
\caption{Top: Mid-infrared observations compared to synthetic spectra
  for three models with the tropopause at 0.001 bars (blue), 0.01 bars
  (purple), \& 0.1 bars (green); these models illustrate the range of
  temperature/composition possibilities consistent with the data.  We
  also show the best fitting model without a temperature inversion
  (red).  Bottom: Model residuals compared to the mid-infrared
  observations.  The model without a tropopause has the largest
  residuals.  The contribution functions and temperature profiles
  associated with these models are shown in Fig. 5.  The mid-infrared
  photometry are from Knutson et al. 2008 and the mid-infrared
  spectroscopy are from Swain et al. 2008a.
\label{fig:resid}}
\end{center}
\end{figure}

\begin{figure}[h!]
\begin{center}
\epsscale{0.5}
\includegraphics[angle=0,scale=1.0]{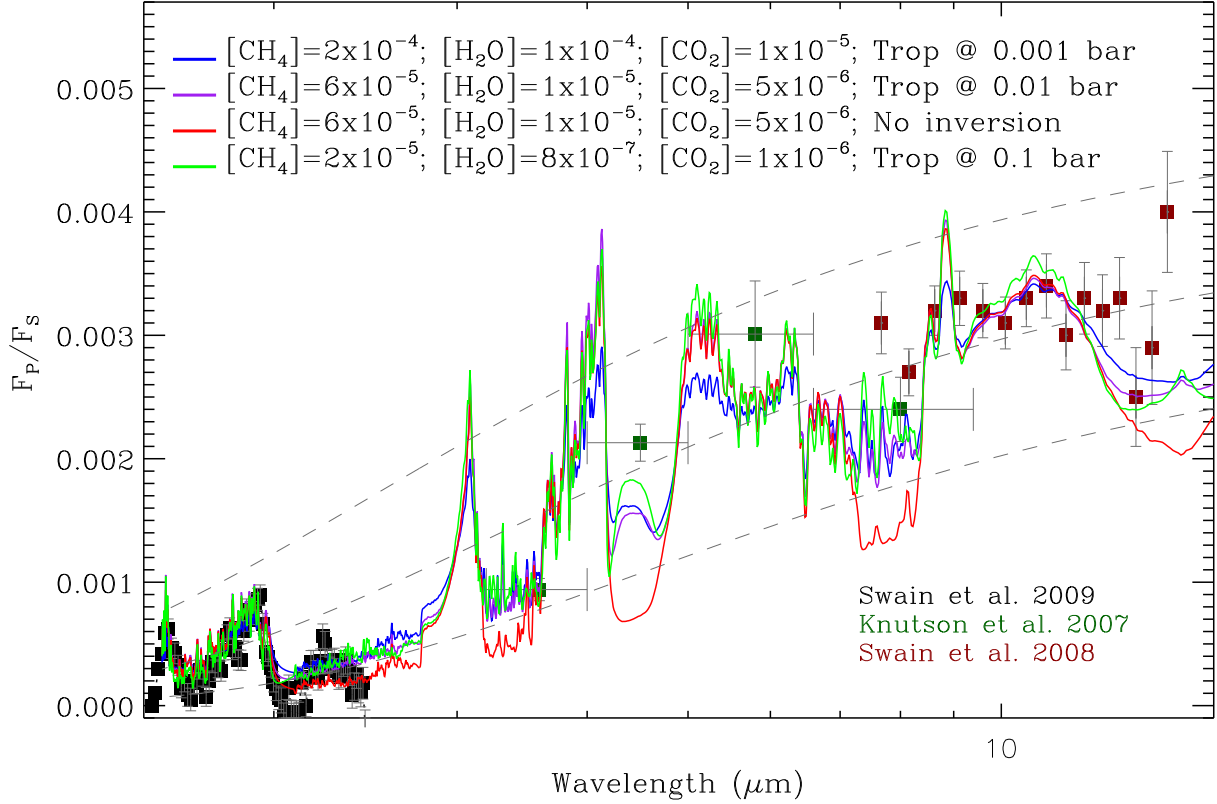}
\caption{The near-infrared and mid-infrared observations compared to
  synthetic spectra for three models (also shown in Fig. 3) that
  illustrate the range of temperature/composition possibilities
  consistent with the data.  For each model case, the molecular
  abundance of CH$_{4}$, H$_{2}$O, \& CO$_{2}$ and the location of the
  tropopause is given; the contribution functions for each of these
  models is shown in Fig. 5.  The three dashed lines correspond to
  single-temperature thermal emission models with temperatures of
  1400, 1800, and 2200 K; these serve to illustrate how the
  combination of molecular opacities and the temperature structure
  cause significant departures from a purely single-temperature
  thermal emission spectrum.  Note that the mid-infrared data
  are not contemporaneous with the near-infrared data, and attempting
  to ``connect'' these data sets with a model spectrum is potentially
  problematic if significant variability is present.}
\end{center}
\end{figure}

\begin{figure}[h!begin{figure}[h!]
\begin{center}
\epsscale{0.5}
\includegraphics[angle=0,scale=1.1,width=8.0cm]{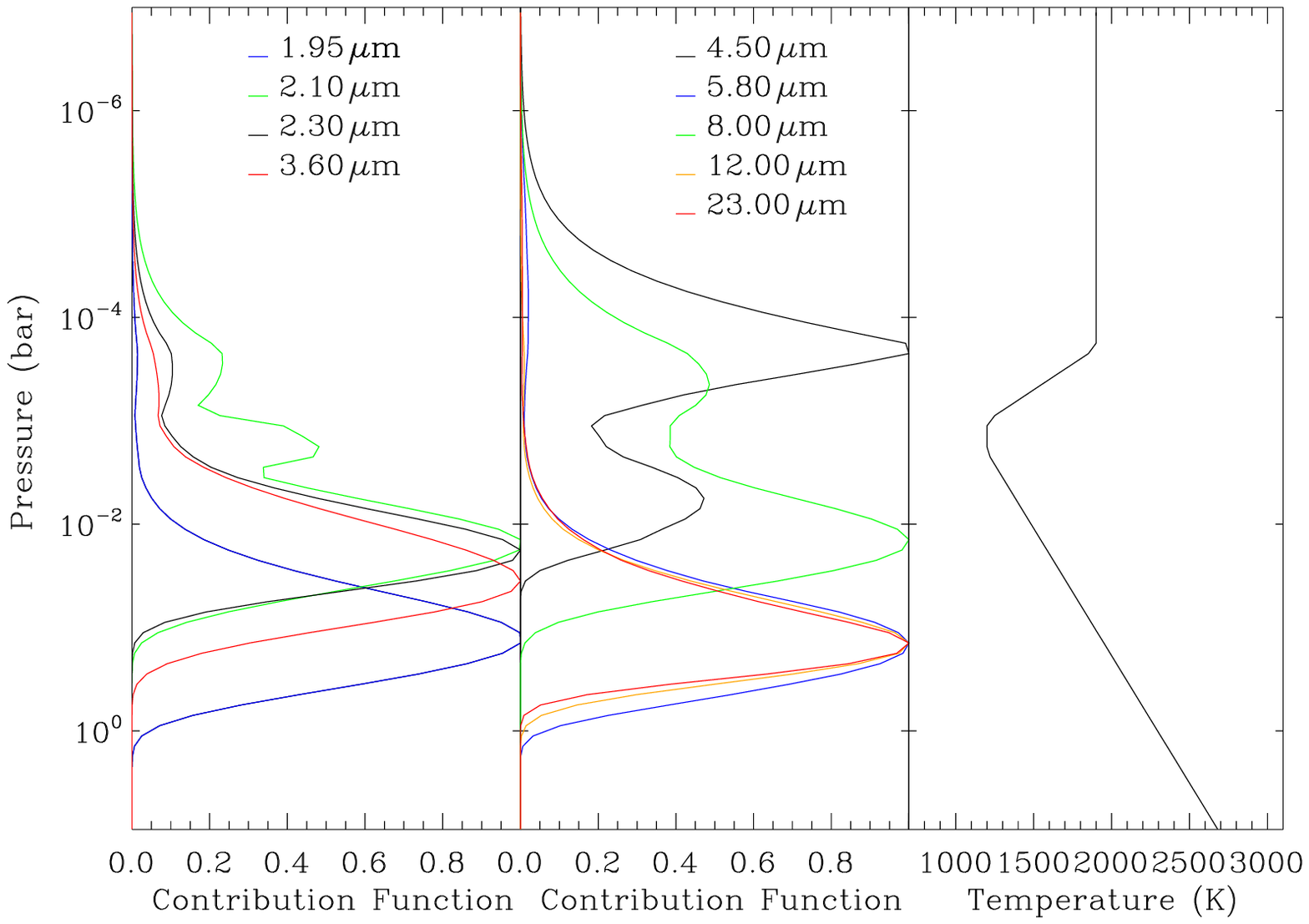}
\includegraphics[angle=0,scale=1.1,width=8.0cm]{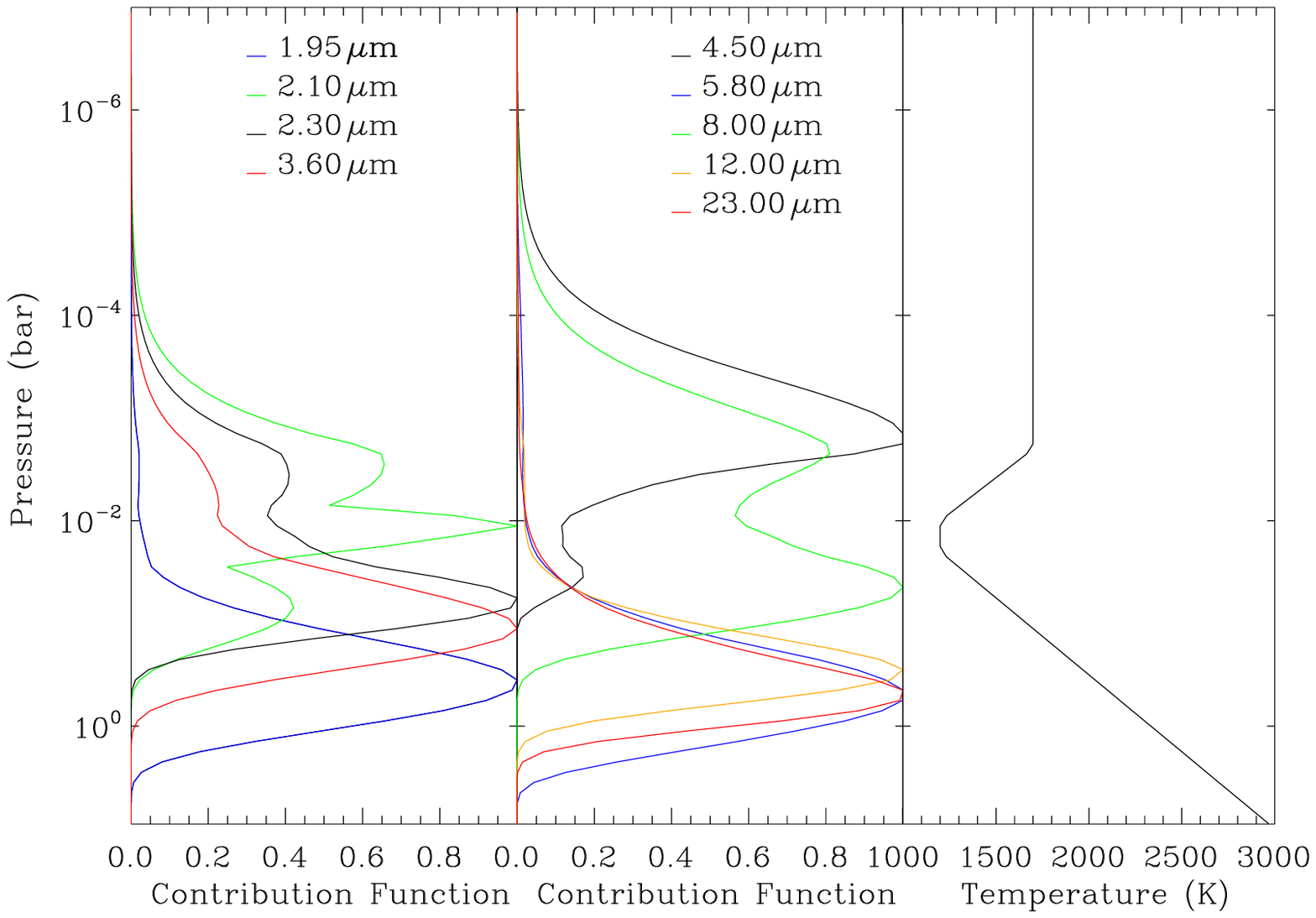}
\includegraphics[angle=0,scale=1.1,width=8.0cm]{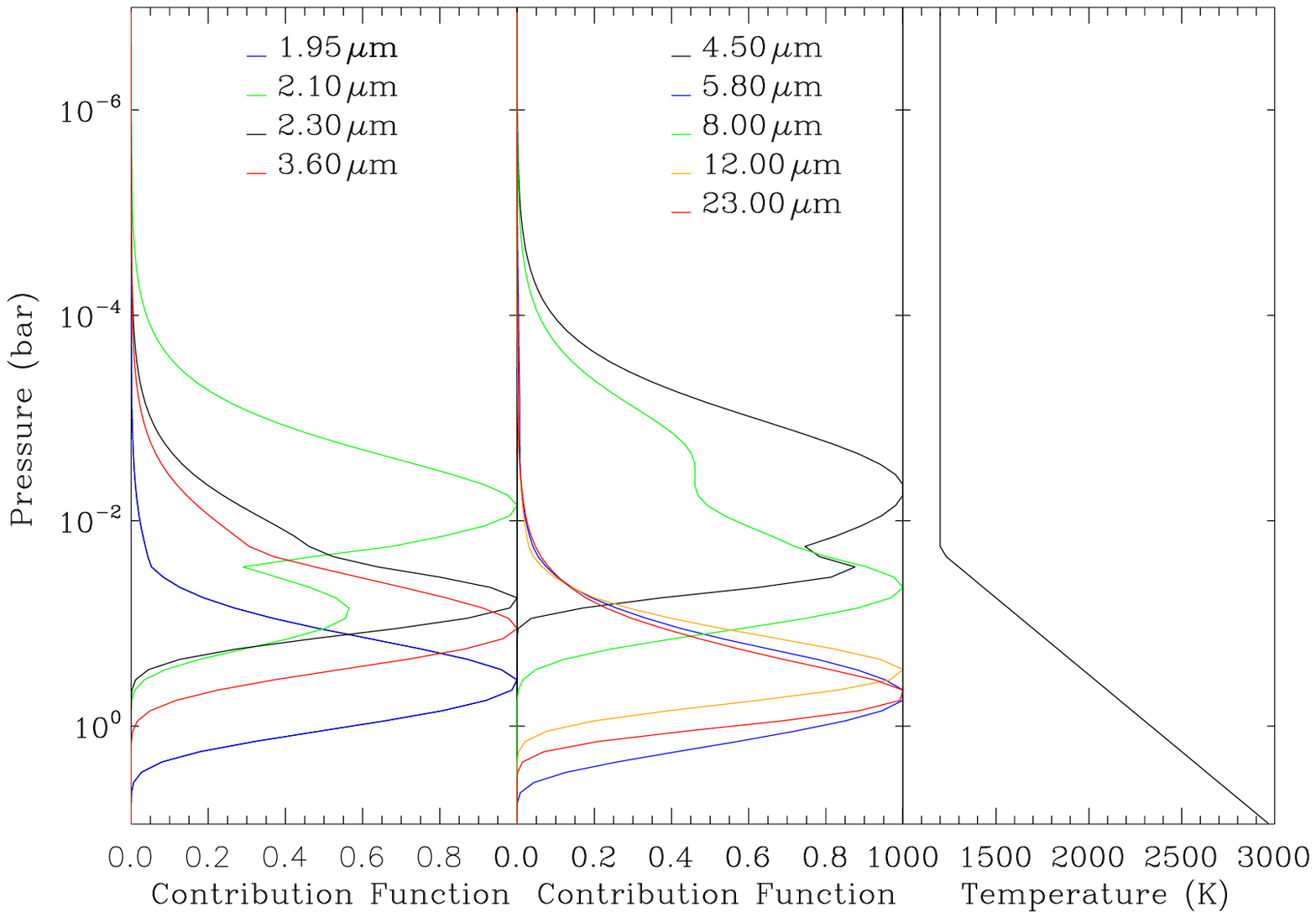}
\includegraphics[angle=0,scale=1.1,width=8.0cm]{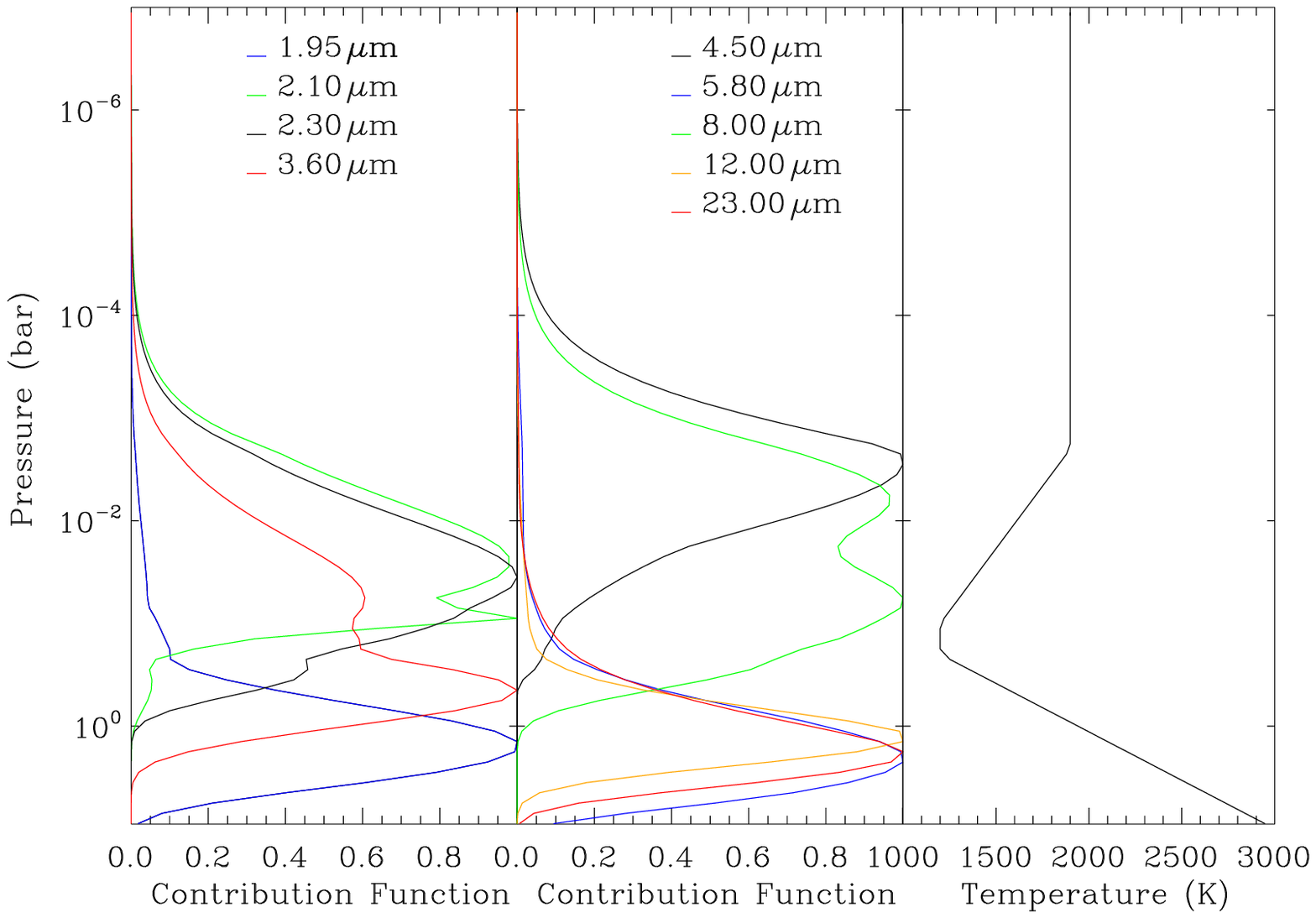}
\caption{The contribution functions for selected wavelengths and the
  associated temperature profile for each model shown in Figs. 3 \& 4;
  starting from the top left and going clockwise, the models correspond
  to a tropopause at 0.001 bar, 0.01 bar, 0.1 bar, and no tropopause,
  respectively.  The selected wavelengths correspond to bands of
  H$_{2}$O and CH$_{4}$; the contribution function is determined by
  opacity in each layer, which, in turn, can depend on the local
  temperature.}
\end{center}
\end{figure}

\begin{figure}[h!begin{figure}[h!]
\begin{center}
\epsscale{0.5}
\includegraphics[angle=0,scale=0.55]{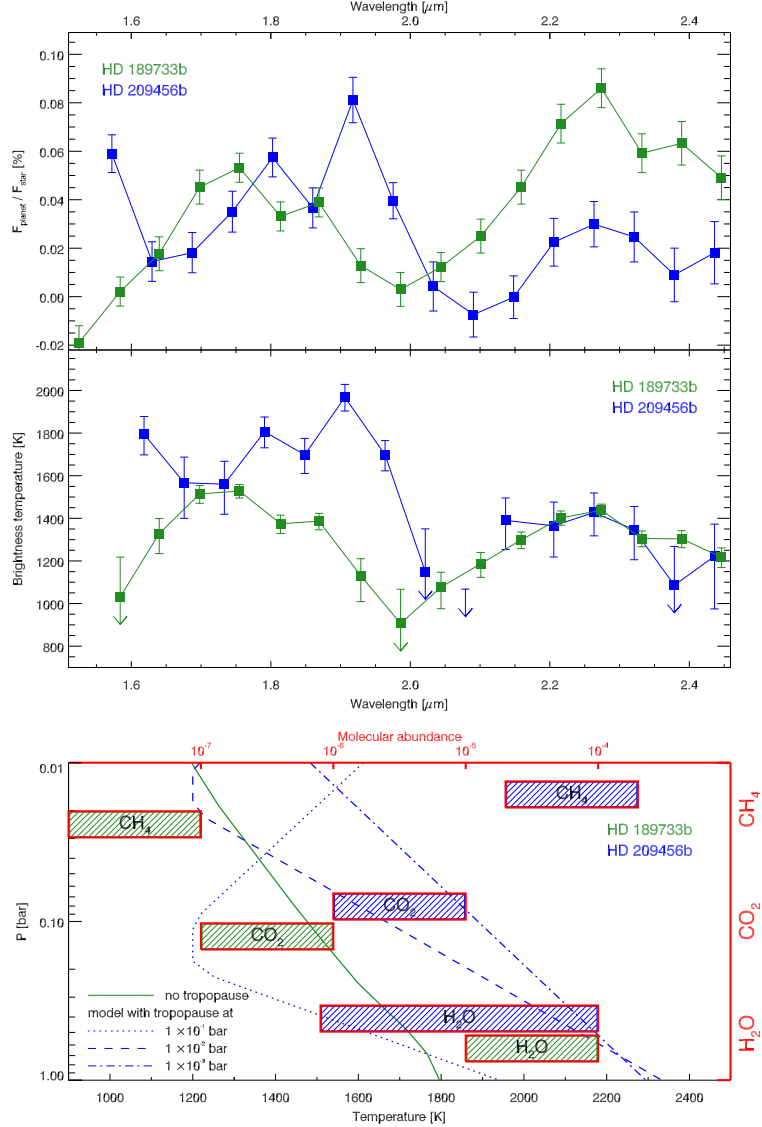}
\caption{Top: The near-infrared dayside emission spectrum of HD~189733b
and HD~209458b, showing the significant differences in the observed
spectra.  Middle: The brightness temperature spectrum of HD~189733b and
HD~209458b.  The spectra probe similar pressure scales in the dayside
atmosphere.  The difference in these spectra is due primarily to the
presence of (1) significant CH$_{4}$ enhancement, and (2) increased
temperature in HD~209458b relative to HD~189733b.  Bottom: A
preliminary comparison of HD~189733b and HD~209458b in terms of the
temperature and molecular abundances.  The abundance ranges shown here
represent the range of solutions we currently identify
as consistent with the data.  Other temperature/abundance solutions
may exist, and thus the results here should be interpreted as
indicative.  There is a suggestion of enhancement in the abundance of
CH$_{4}$ and H$_{2}$O in HD~209458b relative to HD~189733b; this
apparent enhancement needs to be confirmed via additional observations
and detailed modeling.}
\end{center}
\end{figure}


\begin{thebibliography}

\bibitem[Barber et al. (2006)]{barber06} Barber, R.~J., Tennyson, J.,
  Harris, G.~J. \& Tolche\ nov, R.~N. 2006, \mnras, 368, 1087

\bibitem[Burrows, et al. (2007)]{burrows07} Burrows, A., Hubeny, I,
Budaj, J., Knutson, H.~A. \& Charbonneau, D. 2007, \apj, 668, 171

\bibitem[Charbonneau et al. (2000)]{charbonneau00} Charbonneau, D.,
Brown, T.~M., Latham, D.~W., \& Mayor, M. 2000, \apj, 529, 45

\bibitem[Charbonneau et al. (2002)]{charbonneau02} Chabonneau, D.,
Brown, T.~M., Noyes, R.~W., \& Gilliland, R.~L. 2002, \apj, 568, 377

\bibitem[Cowan, et al. (2007)]{cowan07} Cowan N.~B., Agol, E., \&
Charbonneau, D. 2007, \mnras, 379, 641

\bibitem[Deming et al. (2005)]{deming05} Deming, D., Seager, S.,
Richardson, L.~J., \& Harrington, J. 2005, \nat, 434, 740

\bibitem[Fortney et al. (2008)]{fortney08} Fortney, J.~J., Marley,
M.~S., Saumon, D., \& Lodders, K. 2008, \apj, 683, 1104

\bibitem[Griffith, et al. (1998)]{griffith98} Griffith, C.~A., Yelle,
  R.~V. \& Marley, M.~S. 1998, Science, 282, 2063

\bibitem[Grillmair et al. (2008)]{grillmair08} Grillmair, C.~J,
Burrows, A., Charbonneau, D., Armus, L, Stauffer, J., Meadows, V., Van
Cleve, J., von Braun, K., \& Levine, D. 2008, \nat, 456, 767

\bibitem[Knutson et al. (2007)]{knutson07} Knutson, H.~A.,
Charbonneau, D., Noyes, R.~W., Brown, T.~M., \& Gilliland, R.~L.\
2007, \apj, 655, 564

\bibitem[Knutson et al. (2008)]{knutson08} Knutson, H.~A.,
Charbonneau, D., Allen, L.~E., Burrows, A., \& Thomas, M.~S. 2008,
\apj, 673, 526

\bibitem[Nassar \& Bernath (2003)]{nassar03} Nassar, R., \& Bernath,
  P. 2003, JQRST, 82, 279

\bibitem[Rauscher et al. (2007)]{rauscher07} Rauscher, E. et
al. 2007, \apj, 664, 1199

\bibitem[Rauscher et al. (2008)]{rauscher08} Rauscher, E., Kristen,
M., Cho, J.~Y.-K, Seager, S., Hansen, B.~M.S. 2008, \apj, 681, 1646

\bibitem[Rotham et al. (2005)]{rotham05} Rotham, L.~S. et al. 2005,
JQSRT, 92, 139

\bibitem[Rowe et al. (2008)]{rowe08} Rowe, J.~F. 2008,
arXiv:0711.4111v2

\bibitem[Richardson et al. (2007)]{richardson07} Richardson, J.~L.,
Deming, D., Horning, K, Seager, S., \& Harrington, J. 2007, \nat, 445,
892

\bibitem[Showman et al. (2008)]{showman08} Showman, A.~P., Cooper,
C.~S., Fortney, J.~J., \& Marley, M.~S. 2008, \apj, 682, 559S

\bibitem[Swain et al. (2008a)]{swain08a} Swain, M.~R., Bouwman, J.,
Akeson, R.~L., Lawler, S., \& Beichman, C.~A. 2008a, \apj, 674, 482

\bibitem[Swain et al. (2008b)]{swain08b} Swain, M. R., Vasisht, G. V.,
\& Tinetti, G. 2008b, \nat, 452, 329

\bibitem[Swain et al. (2009)]{swain09} Swain M.~R. et al. 2009, \apj,
  960, L114

\bibitem[Tashkun et al. (2003)]{tashkun03} Tashkun, S.~A., Perevalov,
  V.~I., Tefflo, J-L., Bykov, A.~D., Lavrentieva, N.~N. 2003, JQSRT,
  82, 165

\bibitem[Tinetti et al. (2007a)]{tinetti07a} Tinetti, G. et al. 2007a,
\apj, L99

\bibitem[Tinetti et al. (2007b)]{tinetti07b} Tinetti, G. et al. 2007b,
\nat, 448, 169

\bibitem[Walker et al. (2003)]{walker03} Walker, G. et al. 2003,
\pasp, 115, 1023

\bibitem[Zobov et al. (2008)]{zobov08} Zobov, N.~F. et al. 2008,
  \mnras, 387, 1093

\end{thebibliography}
\end{document}